\documentstyle[12pt]{article}

\input{psfig}

\newcommand{\eabe} {\begin{eqnarray}}
\newcommand{\eaen} {\end{eqnarray}}
\newcommand{\eqbe} {\begin{equation}}
\newcommand{\eqen} {\end{equation}}
\renewcommand{\mit} {\mathit}
\newcommand{\mrm} {\mathrm}
\newcommand{\ol} {\overline}

\newcommand{\bibl}[5]
	{#1, {\it #2} {\bf #3} (#4) #5}
\newcommand{\anti}[1] {${\mrm \ol #1 }$}
\newcommand{\pair}[1] {${\mrm #1 \ol #1 }$}

\begin{document}

\begin{titlepage}
\begin{flushright}
  LU TP 96-20\\
  June 1996
\end{flushright}
\vspace{25mm}
\begin{center}
  \Large
  {\bf Baryon Production in the String Fragmentation Picture} \\
  \vspace{12mm}
  \normalsize
  Patrik Ed\'en\footnote{e-mail patrik@thep.lu.se}, G\"osta Gustafson\footnote{e-mail gosta@thep.lu.se}\\
  Department of Theoretical Physics\\
  University of Lund\\
\end{center}
\vspace{6cm}
{\bf Abstract:} \\
An improved version of the ``pop-corn'' model for baryon production in quark and gluon jets is presented. With a reduced number of parameters the model reproduces well both production rates for different baryon species and baryon momentum distributions. Predictions are presented for a set of baryon-antibaryon correlations.
\end{titlepage}

\section{Introduction} \label{sec-intro}
The string fragmentation model~\cite{jet} is able to give a generally very good description of the distribution of hadrons in quark and gluon jets~\cite{revue}. The abundance of mesons with different flavour and spin can be described with a limited number of parameters, which have a natural interpretation. Thus the production of a \pair{q} pair can be understood as a tunneling phenomenon~\cite{HBN}, which gives a suppression of strange quarks compared to the lighter u- and d-quarks. Also the relation between vector and pseudo-scalar mesons and the suppression of tensor mesons can be understood within the string model.

Baryon production is a more complex process than the production of mesons. \pair{B} correlations in rapidity are in agreement with the assumption that the B and the \anti{B} are produced as neighbours or next to neighbours in a string breakup. The distribution in the angle between the baryon and the thrust axis (i.e.\ the general string direction) in the \pair{B} cms is not spherically symmetric~\cite{ppangle1, ppangle2}. 
Furthermore, recent data from polarized $e^+ e^-$ annihilations show that baryons are more frequent in quark jets than in antiquark jets~\cite{SLD}. 
Thus baryon-antibaryon pairs do not originate from isotropically decaying clusters. Instead the distribution can be understood if the B and the \anti{B} are pulled in opposite directions at the string breakup. The simplest possibility would be that the string breaks by the production of a diquark-antidiquark pair, which become constituents in the baryon and the antibaryon, and a model based on this assumption was presented in ref~\cite{bar}. In particular data on transverse momentum correlations~\cite{ppangle1} do not agree with expectations from this simple model, but indicate that occasionally one ore a few mesons may be produced in between the baryon and the antibaryon along the string. This idea was developed in the so called popcorn model~\cite{pop}. Because of the limited experimental information on baryon production, in particular concerning multi-strange baryons, some simplifying approximations were used when the model was implemented in the JETSET Monte Carlo~\cite{pop,JET}. Thus it was e.g.\ assumed in the MC that at most one meson could be produced between the B and the \anti{B}. Until recently this version of the MC was in fair agreement with most experimental data on baryon production. 

With the advent of new data with very high statistics from LEP it is however now possible to achieve a better understanding of the baryon production mechanism. In this paper we will present an improved model for baryon production. This model is based on the same general principles  as the popcorn model in ref~\cite{pop}, but takes better account of the detailed kinematics of the process. The result agrees well with experimental data both for production rates for different baryons and for energy spectra. We also show here some predictions for a set of different correlations, which can be used to test the model.

We want to mention that also other approaches to jet fragmentation have been presented, in particular where the production rates are mainly determined by the different hadron masses. This is the case for the UCLA model~\cite{UCLA}, which is based on the Lund string fragmentation scheme, but in which the production rates are determined by the hadron masses rather than by the quark properties. This model gives a fair description of meson production with few parameters. However, it is not trivial to get a unique prediction for baryon production within this approach, and the most recent publication concentrates on meson production. Chliapnikov et al.~\cite{Chl} have observed regularities between production rates and masses, spin and isospin for different hadrons. However, in this analysis the production of e.g.\ protons and $\Lambda$ include decay products from heavier resonances, and it is not clear to us whether it is possible to find a similar relation for the directly produced hadrons. Also the cluster fragmentation model as implemented in the Herwig MC~\cite{her}, in which hadron rates are mainly determined by their masses, has difficulties to reproduce the experimental data. This further supports our assumption that dynamical properties deeper than just the hadron masses are important in the production of baryons.

In section 2 we review the ideas of the Lund string fragmentation model and of the popcorn model, and also discuss its shortcomings within the default version. In section 3 we describe our revised model, and in section 4 we present some comparisons with experimental data.

\section{String Fragmentation and the Popcorn Model}
\subsection{String fragmentation}

\begin{figure}[p]
  \hbox{
     \vbox{
	\begin{center}
	\mbox{
	\psfig{figure=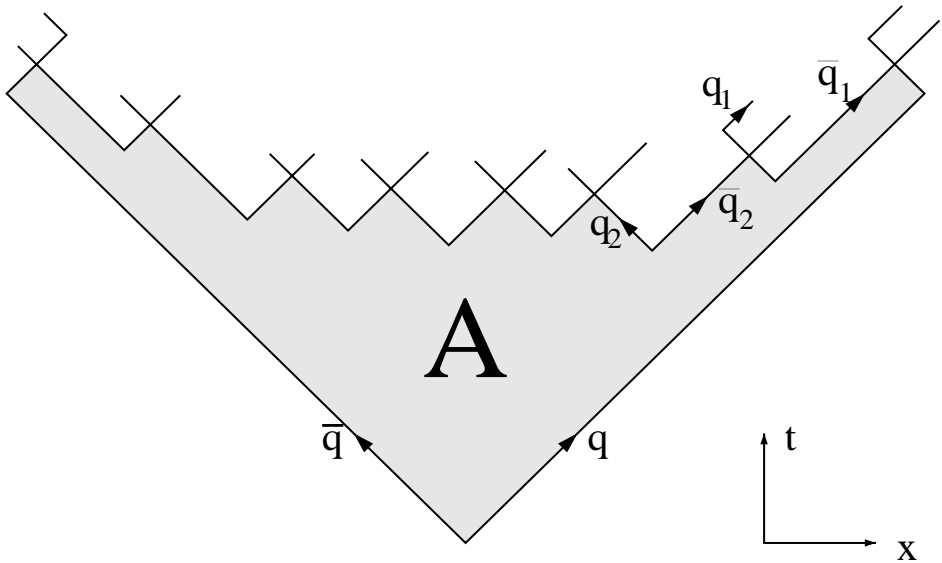,width=10cm,height=14cm}
	}
	\end{center}
    }
  }
  \vspace{-6 cm}
  \caption{\em String fragmentation in x-t space.}
  \label{f:break}
\end{figure}

{\bf Longitudinal momentum distribution}\\
In the string model the confining field is assumed to behave like a relativistic string, i.e.\ like a vortex line in a superconductor. The model contains first a description of the decay of a straight string, and secondly the assumption that gluons behave as transverse excitations or kinks on the string. As gluon kinks have no inherent influence on particle composition, we will in the following discuss the fragmentation of a simple straight string. The string can break by the production of \pair{q} pairs, which are pulled apart by the string tension. When a quark meets an antiquark from a neighbouring pair, they can form a final state meson, as shown in Fig~\ref{f:break}. The production points lie around a hyperbola in $x$-$t$ space, which means that on the average the mesons are evenly distributed in rapidity. When a \pair{q} pair is produced, the system is split in two causally disconnected pieces, and assuming that these pieces decay independently, one finds a unique solution~\cite{sym}. The probability to find a definite final state with meson momenta $p_i$ is given by the following relation, if we assume that there is only one meson species with mass $m$,
\eqbe \mrm Prob. \propto \prod_{\mit i} ({\mit N}d^2{\mit p_i\delta ( p_i}^2-{\mit m}^2))\mit \delta (\sum_k p_k-p_{\mrm tot})\exp(-bA\kappa ^{\mrm 2}). \label{e:Ptot} \eqen
Here $A$ is the (colour coherence) area indicated in Fig~\ref{f:break}, $\kappa$ is the string tension and $N$ and $b$ are two free parameters. This expression has the form of a phase space factor (determined by the parameter $N$) times an exponential term which has the form or a Wilson area law (with the scale parameter $b$). The distribution in Eq~(\ref{e:Ptot}) can be generated iteratively from one end of the meson system in Fig~\ref{f:break}. Thus, if a set of mesons are generated and each one takes a fraction $z$ of the remaining light cone momentum $E+p$ (or $E-p$ if the system is generated from the other end) then $z$ is given by the following distribution
\eqbe f(z)=N\frac{(1-z)^a}{z}\exp(-bm^2/z). \label{e:fz_one} \eqen
The parameters $N$, $b$ and $a$ are related by normalization, leaving two free parameters.
With different quark and hadron species a larger freedom is consistent with the assumptions given above. Different quark flavours can be produced at different average proper times. For a meson with transverse mass $m_\perp$, built up by flavours $\alpha$ and $\beta$, the distribution in Eq~(\ref{e:fz_one}) is then replaced by the expression
\eqbe f_{\alpha \beta}(z) \propto \frac{1}{z}z^{a_\alpha}\left( \frac{1-z}{z} \right )^{a_\beta}\exp(-bm_\perp^2/z). \label{e:fz} \eqen 
Here $\alpha$ is the flavour of the previously produced \pair{q} pair, and $\beta$ the flavour of the new \pair{q} pair. The proper time, $\tau$, of the break-up points, or the variable $\Gamma$ defined by
\eqbe \Gamma \equiv \kappa^2 \tau^2, \label{e:Gdef} \eqen is given by the probability distribution
\eqbe H(\Gamma) \propto \Gamma^{a_\alpha} e^{-b\Gamma}. \label{e:HG} \eqen 
From this relation we get the mean value \eqbe \left< \Gamma_\alpha \right> = \frac{1+a_\alpha}{b}. \label{e:meanG} \eqen
Thus with different parameters $a_\alpha$, the proper time for the breakup points may depend on the flavour $\alpha$ of the \pair{q} pair produced. In fits to experimental data different $a$-values have usually only been introduced for effective diquarks in connection with baryon production.

{\bf Flavour and transverse momentum}\\
Quarks with mass $\mu$ and transverse momentum $k_\perp$ must classically be produced in the string at a certain distance, $d$, so that the string energy between them can be transformed into the transverse mass of the pair $2\mu_\perp = 2\sqrt{\mu^2+k_\perp^2} = \kappa d$. Quantum mechanically this production can be described as a tunneling process~\cite{HBN,jet}. The wave functions of the quark and the antiquark are exponentially damped in the classically forbidden region, and the production probability is proportional to the square of the product of these wave functions. Using the WKB approximation we find that this is approximately given by
\eqbe \left| \psi_{\mrm \ol q}\psi_{\mrm q} \right|^2 \sim \exp(-\frac{\pi}{\kappa}\mu^2_\perp)=\exp(-\frac{\pi}{\kappa}\mu^2)\exp(-\frac{\pi}{\kappa}k^2_\perp) \label{e:M} \eqen
if the wave functions $\psi$ are normalized to 1 at the classical boundary. This expression factorizes in a flavour factor (s-quarks will be suppressed by about 1/3, while charm and heavier quarks can be neglected) and a Gaussian $k_\perp$ distribution.

\begin{figure}[tb]
  \hbox{
     \vbox{
	\begin{center}
	\mbox{
	\psfig{figure=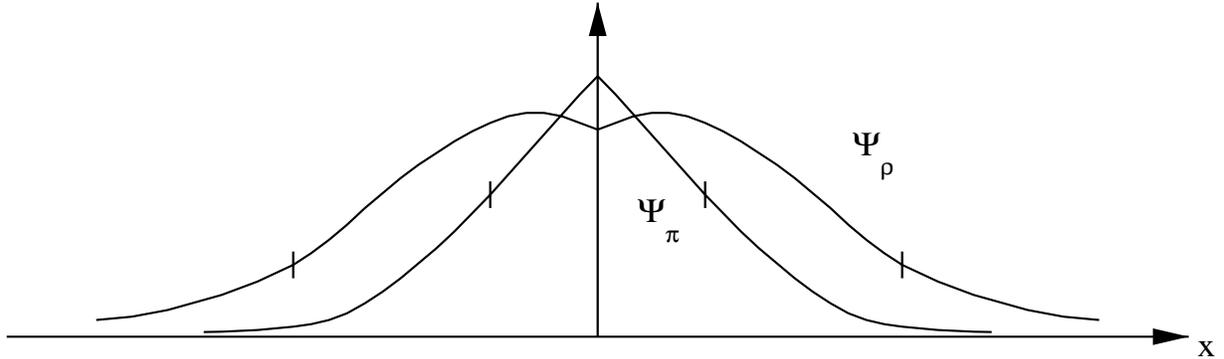,width=7.5cm,height=10cm}
	}
	\end{center}
    }
  }
  \vspace{-4 cm}
  \caption{\em Wave functions for $\rho$ and $\pi$, showing the different normalization at the classical turning point, which is marked with a dash.}
  \label{f:phi_c}
\end{figure}

{\bf Vector meson suppression}\\
The result in Eq~(\ref{e:M}) was obtained if the wave functions were normalized to 1 at the classical boundary. The produced quark must however fit into the wave function of a final state meson. A spin-spin interaction implies e.g.\ that the wave function for a $\rho$-meson is more spread out then that of a pion, see Fig~\ref{f:phi_c}. Thus the wave function at the boundary is smaller for $\rho$, and it has been estimated that~\cite{jet}
\eqbe \psi^2{\rm (class.\ boundary)} \sim 1/m_\perp \label{e:spinsupp} \eqen
where $m_\perp$ is the transverse mass of the meson.

In this way we can understand why the $\rho/\pi$ ratio is suppressed from 3:1, as obtained from spin counting, to something closer to 1:1. The $p_\perp$ dependence in Eq~(\ref{e:spinsupp}) is weak for most mesons except for the pions, which have very small mass. For pions we expect from Eq~(\ref{e:spinsupp}) an enhancement for small $p_\perp$ values and a more narrow $p_\perp$ distribution. Due to the small pion mass we also expect further correlations in $p_\perp$ between neighbouring pions, see further ref~\cite{jim}.

\subsection{Baryon production}
As mentioned in the introduction the simplest model for baryon production would follow from the assumption that the string can break by the production of a diquark-antidiquark pair in a \anti{3}-3 colour state. An antidiquark with specified flavour and spin would then appear as a new quark flavour in Eqs~(\ref{e:fz},\ref{e:HG},\ref{e:M}), and different production rates would be determined by effective diquark masses and spin suppression factors. (Because baryons are symmetric three-quark states, extra weights have also to be applied in order to guarantee SU(6) symmetry.) As the effective masses are difficult to determine theoretically, they were represented by a set of free parameters in the MC implementation. These parameters determine the relative production ratios ${\rm qq/q}$, $\mrm ud_1 /ud_0$, $\mrm us/ud$. The MC also includes a possibility to further suppress the decuplet baryon states relative to the octet states, but the corresponding parameter is usually set equal to one.

\begin{figure}[tb]
  \hbox{
     \vbox{
	\begin{center}
	\mbox{
	\psfig{figure=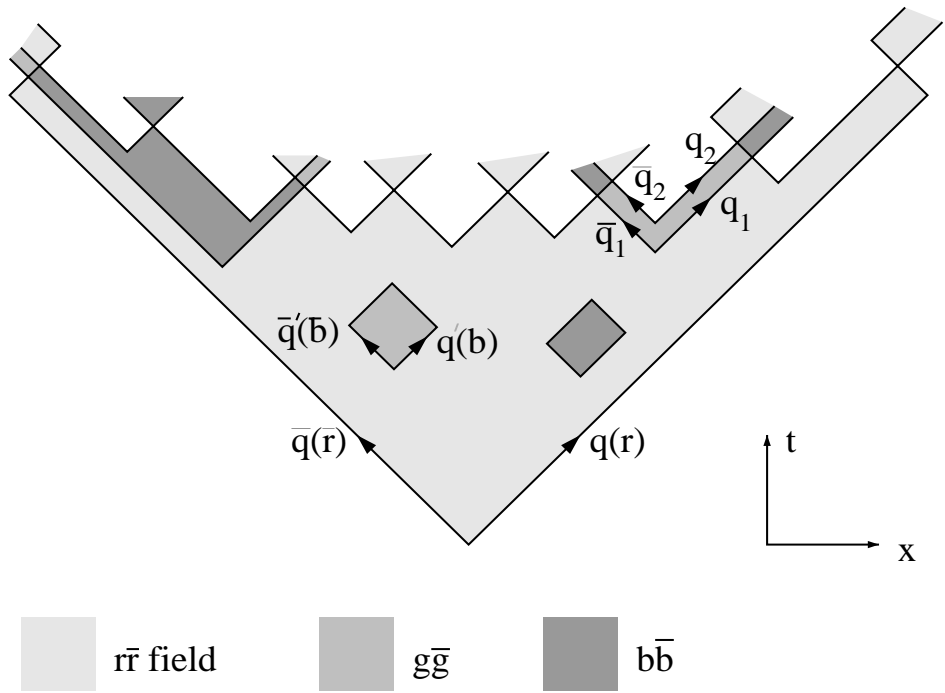,width=10cm,height=14cm}
	}
	\end{center}
    }
  }
  \vspace{-5 cm}
  \caption{\em If a vertex pair \pair{q_2} is produced inside a colour fluctuation region spanned by \pair{q_1}, an effective diquark-antidiquark pair has ``popped out'' in a stepwise manner. The model allows for several breakups in the colour fluctuation region, creating one or several mesons in between the baryon and antibaryon, as shown to the left. }
  \label{f:pop}
\end{figure}

To improve agreement with experiments a more general framework for baryon production was presented in the ``popcorn'' model~\cite{pop}. In this model a diquark is produced in a stepwise manner and not as a single unit. Assume that a colour field is stretched between e.g.\ a red quark and an antired antiquark. The string can break if a \pair{r} \pair{q} pair is produced and pulled apart by the string tension. We can also imagine that a \pair{b} pair is produced as a virtual fluctuation. If the rb (\anti{r}\anti{b}) is in a colour antitriplet \anti{g} (triplet g) state, the colour field between the produced quark and antiquark will correspond to a triplet colour field with the same strength as the original field. Thus equal forces in opposite directions act on the new quark and the new antiquark,  leaving no net force on either of them. In accordance with the uncertainty principle they can move around freely for a time inversely proportional to their energy. These quarks are in the present paper called ``curtain quarks'', from the picture of rings on a curtain-rod, sliding back and forth with no frictional losses and without changing the properties of the rod. In distinction to curtain quarks, the quarks that cause the string to break are in this paper called ``vertex quarks''. If the string breaks within the colour fluctuation region, an effective diquark-antidiquark pair is produced, see Fig~\ref{f:pop}.

As no net force is acting on the curtain quarks they appear as free particles. Therefore it was in ref~\cite{pop} expected that the probability to find such a \pair{q} pair with transverse mass $\mu_\perp$ at a separation $d$ in space is proportional to $\left|\Delta_F(d,\mu_\perp)\right|^2$. To produce an intermediate meson system with invariant mass $M$ between the baryon and antibaryon, the distance $d$ must be larger than $M/\kappa$. As $\Delta_F$ falls off exponentially for large $d$-values, a strong suppression is obtained for large masses $M$. For this reason the possibility to have more than one intermediate meson was neglected in the MC implementation. The probability for one intermediate meson was in the program determined by the so called popcorn parameter. In the popcorn model the production ratios between different baryon species become slightly modified since a specific diquark is not always accompanied by a corresponding antidiquark. Thus $\Omega$ production is somewhat less suppressed, as the antiquark partners of the ss diquark not necessarily go together in a \anti{s}\anti{s} antidiquark.

Turning to the momentum distribution we note that baryon spectra in the default JETSET program is harder than the experimental results~[15-19]. In the popcorn model the two quarks in an effective diquark are produced at different proper times (cf Fig~\ref{f:pop}). The relevant time in Eqs~(\ref{e:Gdef}) and (\ref{e:meanG}) is the latest one, the one corresponding to the ``vertex quark'', and it is natural that on average this is larger than the time for the \pair{q} production associated with meson production outside the colour fluctuation region. Thus we could expect a larger value of $a$ for the effective diquark production (cf Eq~(\ref{e:meanG})), which according to Eq~(\ref{e:fz}) results in a softer baryon spectrum. This effect is however not large enough, and in attempts to describe experimental data it has been suggested that first rank baryons might be suppressed~\cite{tune}. (Rank denotes the order in which the hadrons are produced, counting from the string end. Thus in Fig~\ref{f:break}, the meson built up by the original quark q and the antiquark \anti{q_1} has rank 1, the meson build up by ${\rm q_1}$ and \anti{q_2} has rank 2 etc.) In the string model heavy baryons like $\mrm \Lambda_c$ or $\mrm \Lambda_b$ are only created as first rank particles (or decay products of these), and in order to get enough of these heavy baryons, the suppression of the first rank baryons should not apply to ${\rm c}$ or ${\rm b}$ jets. If the curtain quarks move along the light-cones as indicated in Fig~\ref{f:pop}, a special suppression of first rank baryons may not appear natural, but, as we will show in the following section, it can indeed be dynamically understood in a more detailed analysis of the kinematics.

\section{The Revised Model} \label{sec-model}
\subsection{Baryon flavours} \label{sec-model flavours}

In the revised model we take the separate production of the curtain quarks and the vertex quarks more seriously. Thus we assume that the production of a vertex \pair{q} pair, which breaks the string, is always determined by the expression in Eqs~(\ref{e:M}), irrespective of whether it is ending up in an effective diquark or not. The s-quark suppression of all {\em vertex} quarks is therefore essentially determined by the ${\rm K}/\pi$ ratio.

Turning to curtain quarks we assume, following ref~\cite{pop}, that the production probability for these quarks is determined by the function $\Delta_F(d,\mu_\perp)$. Here $\mu_\perp$ is the transverse mass of the curtain quark and $d$ is the minimal separation of the pair. If a meson system with invariant (transverse) mass $M$ is produced between the baryon and the antibaryon, $d$ is given by the sum of $M$ and the effective diquark masses (called $m_1$ and $m_2$) entering the baryon and the antibaryon. Thus \eqbe \kappa d = M+m_1+m_2. \label{e:popdistance} \eqen For large values of $d$, $\Delta_F$ is exponentially suppressed and as in the previous implementation we use the approximation \eqbe |\Delta_F(d ,\mu_\perp)|^2 \sim \exp(-2\mu_\perp d). \label{e:pop} \eqen The relative production rates for different diquarks are thus governed by their mass differences. Even though diquark masses must be considered unknown, we assume that the differences in mass can be determined from the constituent masses, as derived in the parton model.

As the quantity $\mu_\perp$ in Eq~(\ref{e:pop}) is the transverse mass of the curtain quark, we note that large transverse momenta are strongly suppressed. Thus $k_\perp$ for the curtain quarks are generally small compared to the $k_\perp$ of the vertex quarks, and has therefore been neglected in the MC implementation. Consequently the transverse momentum of the hadron will essentially be given by two vertex quarks, and -- like the earlier versions -- the new model predicts a similar $p_\perp$ distribution for primary baryons and mesons~\footnote{Neglecting the rest mass of the quarks and using a typical value for $\kappa d \approx 1 \mrm GeV$, we find $P(k_\perp)\propto \exp(-2d k_\perp) \Rightarrow \left< k_\perp \right> \sim 0.1 \mrm GeV \label{e:kt}$. As the $k_\perp$-contributions to a hadron are essentially added in quadrature, this is very small compared to the transverse momenta of the vertex quarks, for which we have $\left<k_\perp \right>_{\mrm (vertex)}\sim 0.4 \mrm GeV$.}. This approximation should be very good for inclusive distributions, but may be somewhat too crude for correlations in $\bar p_\perp$ or azimuthal angle. In the implementation, we let two parameters \eqbe \beta_q \equiv 2\left<\mu_{\perp \mrm q}\right>/\kappa,~~{\mrm or}~~\beta_{\mrm u}~{\mrm and}~\Delta\beta \equiv \beta_{\mrm s}-\beta_{\mrm u}, \label{e:betadef} \eqen represent $\left<\mu_\perp\right>$ in Eq~(\ref{e:pop}) for light and strange curtain quarks.

From Eqs~(\ref{e:pop}) and~(\ref{e:betadef}), the relative production rate for different diquarks with the same type of curtain quark (light or strange) can be expressed by diquark mass differences and two $\beta$ parameters. The ratio between diquarks with different types of curtain quarks depends not only on mass differences but also on $m_{\mrm ud_0}$. Thus e.g. \eabe  \frac{P\mrm(us_0,s)}{P\mrm(ud_0,u)} & = & \exp(-2(\beta_{\mrm s}m_{\mrm us_0}-\beta_{\mrm u}m_{\mrm ud_0}))\nonumber\\
& = & \exp(-2\beta_{\mrm s}(m_{\mrm us_0}-m_{\mrm ud_0}))\exp(-2\Delta\beta m_{\mrm ud_0}) \label{e:mud0effect}. \eaen 
At first this seems to call for a new free parameter. It turns out, however, that by changing the $\beta_s$ value and the decuplet suppression discussed below, one can compensate all effects of a variation in $m_{\mrm ud_0}$, apart from a small variation on the $\Delta$ production rate. In view of the experimental uncertainty regarding $\Delta$ production, we find little need to introduce $m_{\mrm ud_0}$ as a free parameter, and in the MC its value is therefore fixed to $0.2$GeV.

In the same way as for mesons, we also for baryon production expect a spin dependent factor related to the normalization of the hadronic wave function. Also the earlier MC contains one parameter for suppression of the decuplet states compared to the octet states, but in the default version this parameter is set to 1, because the decuplet is sufficiently suppressed by smaller weights for production of the heavier spin 1 diquarks. In our model the three quarks in a baryon are produced essentially independently. Thus we expect a suppression of higher spin states, in a similar way for baryons as for mesons, due to the normalization of the total three-particle wavefunction. Following the arguments in ref~\cite{jet} we assume that this factor is approximately linear in the baryon mass (cf also with the $\pi^0$-$\eta$-$\eta'$ ratios discussed in ref~\cite{jim}). Thus in addition to a suppression factor for the decuplet states compared to the octet states, we also expect a suppression of $\Sigma$ compared to $\Lambda$ due to the larger spin-spin interaction in ud quark pairs compared to us or ds pairs. In the MC program we have introduced one parameter for the decuplet/octet suppression while the extra $\Sigma$/$\Lambda$ suppression from the mass differences ($M_{\Sigma}-M_{\Lambda} \sim 70 {\mrm GeV},~~M_{\mrm decuplet}-M_{\mrm octet} \sim 200\mrm GeV $) is fixed by the decuplet/octet suppression through the relation \eqbe \frac{P\mrm (decuplet)}{P\mrm (octet)}=\frac{1}{1+Q} ~~~\Rightarrow ~~~\frac{P(\Sigma)}{P(\Lambda)}=\frac{1}{1+Q/3}. \eqen This suppression is implemented in such a way that the total production of $\Lambda$ and $\Sigma$ is kept constant.

{\bf Popcorn mechanism}\\
Using the two $\beta$ parameters, Eq~(\ref{e:popdistance}) and ~(\ref{e:pop}), the popcorn mechanism has been implemented without the introduction of any new parameter.

When the popcorn model was implemented in the JETSET MC it was assumed that the production of two or more mesons between the baryon and the antibaryon could be neglected, and there was no extra vector meson suppression in spite of the larger mass compared to the pion mass. This was a deliberate approximation assuming that the overestimated $\rho$ and $\omega$ production could simulate the neglected production of two or three uncorrelated pions. In the new MC this approximation is not used, and the probability for a ``popcorn meson system'' is determined by the invariant mass of this system. In table~\ref{tab:Npop} we show the rate of popcorn systems with different number of popcorn mesons for different values of $\beta_{\mrm u}$. 

\begin{figure}[tb]
  \hbox{
     \vbox{
	\begin{center}
	\mbox{
	\psfig{figure=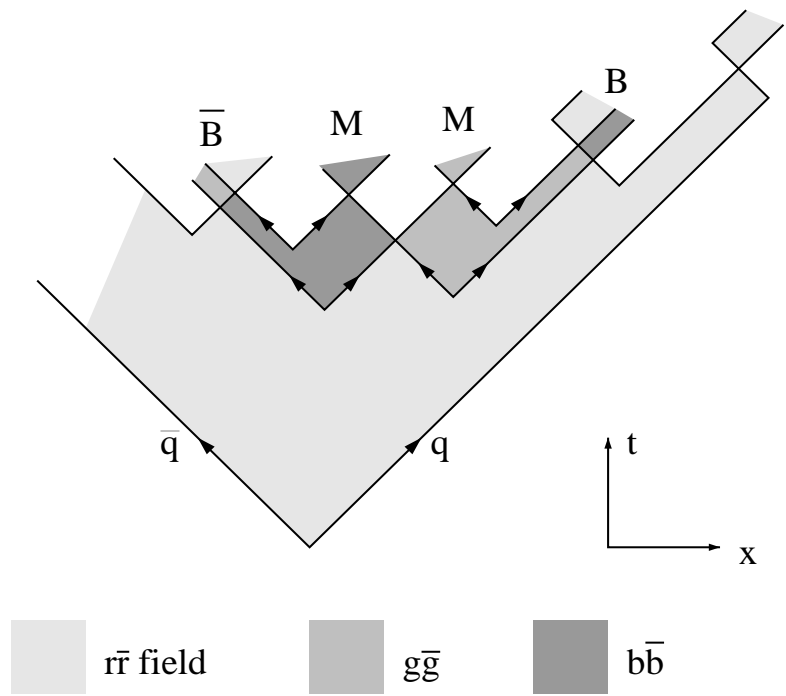,width=10cm,height=14cm}
	}
	\end{center}
    }
  }
  \vspace{-4 cm}
  \caption{\em A popcorn example with two curtain quark pairs.}
  \label{f:DoubleCurtain}
\end{figure}

Popcorn systems with more than one meson could also be formed by two separate colour fluctuations as illustrated in Fig~\ref{f:DoubleCurtain}. In such a situation the baryon and antibaryon need not have any flavour in common. We expect this production mechanism to be suppressed, and it is neglected in the MC. If the production rate for each curtain quark pair is determined by the ratio $P(\mrm qq)$/$P(\mrm q) \sim 0.1$, the ``double curtains'' should be suppressed cf to the ``single curtains'' with multiple popcorn mesons by this factor. As the latter is only a small fraction of the total baryon production (15\% for $\beta_{\mrm u}$, cf tab~\ref{tab:Npop}), we believe the double curtains to give a very small contribution, even if they are somewhat less suppressed by the factors on Eqs~(\ref{e:popdistance}) and~(\ref{e:pop}). In principle this kind of baryon production can be experimentally indicated if one can observe e.g.\ correlated $\Omega^-\mrm \ol p$ or $\Delta^{++} {\ol \Xi}^+$ pairs, which can only be produced by the double curtain mechanism.

Summarizing, apart from the s/u parameter determined by the meson rates and the diquark probability parameter (qq/q), the baryon multiplicities are in our model essentially determined by three tunable parameters, $\beta_u$, $\Delta\beta$ and the decuplet/octet suppression, giving a total of five parameters. This should be compared to the earlier model and its implementation in JETSET, in which all together seven parameters govern baryon production (if the decuplet suppression is not included): The s/u and qq/q parameters plus five more, one suppressing strangeness in effective diquarks, one suppressing spin 1 diquarks, and three related to the popcorn mechanism. 

\subsection{Momentum distributions}
We have assumed that the curtain quarks appear as massive particles. This implies that they cannot move along light-cones as in the illustration in Fig~\ref{f:pop}. Since the curtain quarks behave as free particles we assume that the curtain quark and its antiquark partner move along straight lines from a common starting point. We would then expect a suppression if some of the potential starting points lie outside the color field region. To estimate this effect we will be guided by a simple  semi-classical model. (We use the semi-classical model mainly to estimate the relative suppression when the extension of the colour field is reduced. The overall normalization is determined by the quantum mechanical expression $\Delta_F$ in Eq~(\ref{e:pop}).) As we will see, the net result is first a larger effective $a$-parameter for baryon production (cf Eq~(\ref{e:fz})), second a suppression of first rank baryons. This suppression is stronger for light baryons, but rather weak for baryons containing heavy c- or b-quarks. Thus this model provides a dynamical interpretation of the phenomenological observations in ref~\cite{tune}.

\begin{figure}[tb]
  \hbox{
     \vbox{
	\begin{center}
	\mbox{
	\psfig{figure=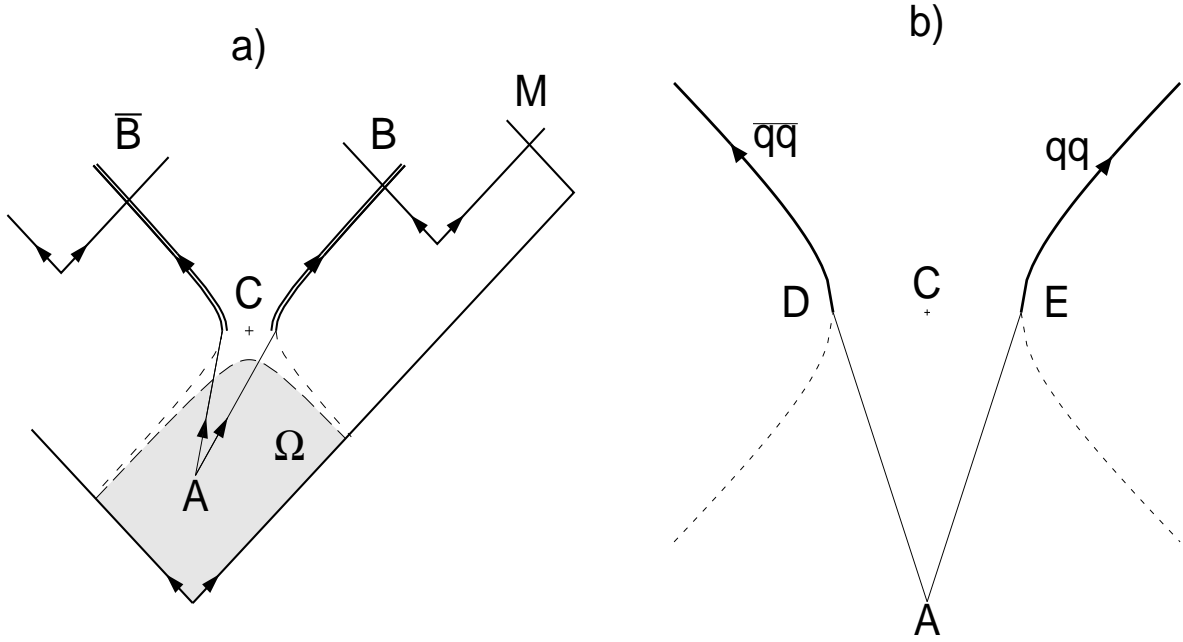,width=10cm,height=14cm}
	}
	\end{center}
    }
  }
  \vspace{-4 cm}
  \caption{\em Curtain pair production at point A, and a diquark vertex at point C. Figure b is boosted to the frame where A and C are at a common space point.}
  \label{f:Pgamma}
\end{figure}

Let us first study a situation with only one vertex quark, i.e.\ a situation with no meson between the baryon and the antibaryon. Assume that a curtain quark-antiquark pair is produced in a point (called $A$ in Fig~\ref{f:Pgamma}a) at an invariant distance $\tau$ from the vertex point $C$ for the effective diquark-antidiquark pair. Fig~\ref{f:Pgamma}b illustrates the situation in a frame where the space distance between $A$ and $C$ is zero. (Consequently the time distance equals $\tau$.) We assume that the curtain quark and antiquark combine with their vertex quark partners in the points $D$ and $E$ to form an effective diquark and antidiquark. This diquark (antidiquark) is not massless, and thus it moves along a hyperbola with centre in the vertex point $C$. If the mass of the diquark-antidiquark pair is 2$m_\perp$ the distance $d$ between $D$ and $E$ is given by $d=2m_\perp/\kappa$ (cf Eq~(\ref{e:popdistance})). In the model we assume that the probability for this process is proportional to $\exp(-S)$, where $S$ is the action. Thus, if $\mu_\perp$ is the transverse mass of the curtain quarks we get a relative weight 
\eabe 
w & = & \exp(-S) \nonumber\\
S & = & E*t    \label{e:actsupp}\\
E & = & 2\mu_\perp/\sqrt{1-v^2}=2\mu_\perp/\sqrt{1-d^2/4t^2}. \nonumber\eaen
We note that the curtain quark and the vertex quark could also combine at a different point along the hyperbola, and similarly for the antiquarks. Also the two recombinations do not have to be at the same time in the particular frame chosen in Fig~\ref{f:Pgamma}b. Other possibilities would generally give smaller weights, and therefore we use for the general situation in Fig~\ref{f:Pgamma}a the weight in Eq~(\ref{e:actsupp}) with $t$ replaced by the proper time $\tau$ between the points $A$ and $C$.

In a situation with more than one vertex \pair{q} pair, we have an intermediate meson system with (transverse) mass $M$. The minimum distance between the ``combination points'' is now given by $d=(M+m_{\perp 1}+m_{\perp 2})/\kappa$ (cf. Eq~(\ref{e:popdistance})), and we will assume that also here the weight for the curtain quark production is given by Eq~(\ref{e:actsupp}) with this value of $d$.

In a situation as illustrated in Fig~\ref{f:Pgamma}a, the total weight $W$ is obtained by summing over all production points $A$ inside the kinematically allowed region $\Omega$, defined by the condition $\tau>m_\perp/\kappa$. \eqbe W=\int_\Omega wdxdt \label{e:intG}. \eqen The result is a total weight $W(\Gamma,\mu_\perp,d)$ which is a function of the three variables $\Gamma$ (the square of the proper time for the effective diquark vertex), $\mu_\perp$ (the transverse mass of the curtain quark) and $d$ (the minimum separation for the curtain quark-antiquark pair).

The effect of this ansatz can be anticipated: The action in Eq~(\ref{e:actsupp}) becomes large both for large $t$ and small $t$ (which implies large velocities) and has a minimum for $t^2=d^2/2$. Thus the most favoured region for the production point $A$ lies at some finite distance from $C$. The result in Eq~(\ref{e:pop}) should correspond to a situation when $\Omega$ is large enough to contain most of this region, while the production probability should be further suppressed for small regions $\Omega$.

A numerical calculation shows that for large vales of $\Gamma$ the weight $W$ scales with $\Gamma$ in the following way \eqbe W \sim \Gamma^{\delta a(\mu_\perp,d)} ~~;~\Gamma~{\rm large}\label{e:Gamscale}. \eqen The power $\delta a(\mu_\perp,d)$ has the effect of shifting the $a$-parameter in Eq~(\ref{e:fz}) and~(\ref{e:HG}). 
The $\delta a$ dependence on $d$ and $\mu_\perp$ implies a unique shift of $a$ for each diquark system. However, the differences are not large, and for relevant values of $\mu_\perp$ and $d$ we find $\delta a \approx 0.5$.
  
\begin{figure}[tb]
  \hbox{
     \vbox{
	\begin{center}
	\mbox{
	\psfig{figure=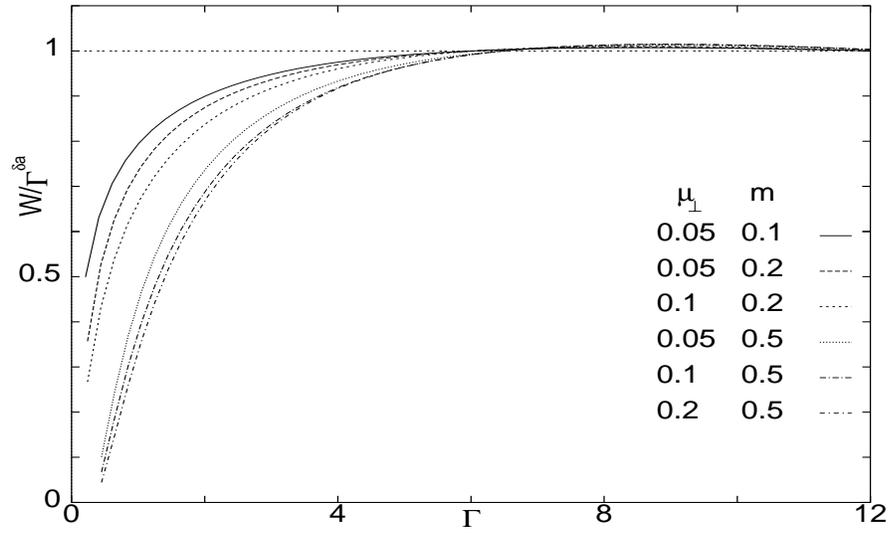,width=8.5cm,height=10cm}
	}
	\end{center}
    }
  }
  \vspace{-3 cm}
  \caption{\em Extra suppression $P(\Gamma)\equiv W/\Gamma ^{\delta a}$ as a function of $\Gamma$ for different values of $m$ and $\mu_\perp$. $W$ is normed so that $P \approx 1 $ for large $\Gamma$.}
  \label{f:Pg}
\end{figure}

The scaling result in Eq~(\ref{e:Gamscale}) is valid only when $\Gamma$ is large compared to $(\kappa d)^2$. The weight $W$ goes to zero when the area of possible production points $\Omega$ is small. As seen from the numerical results shown in Fig~\ref{f:Pg} the extra suppression for small $\Gamma$ differs for different diquark masses and curtain quark masses. From these differences, and the total normalization of $W$ for different $d$ and $\mu_\perp$, production probabilities for diquarks could in principle be estimated within the semi-classical framework. 

In the semiclassical model the production probability goes to zero when \eqbe \Omega=\Gamma-(\kappa d/2)^2\left(1+\ln\frac{\Gamma}{(\kappa d/2)^2}\right) \label{e:Omega} \eqen goes to zero, i.e.\ when $\Gamma$ goes to $(\kappa d/2)^2$. In a proper quantum mechanical treatment the probability should go to zero when $\Gamma \rightarrow 0$. $\Gamma$ values below $(\kappa d/2)^2$ should be suppressed but not totally excluded. Thus we feel that using the exact result in Eq~(\ref{e:intG}) would be to stretch the semiclassical model too far, and instead we use for all values of $\mu_\perp$ and $d$ the following suppression factor for small $\Gamma$-values \eqbe P(\Gamma) = 1 - \exp(-\rho \Gamma)~~\mrm with~~ \rho\approx\ln(2)GeV^{ -2}. \label{e:Gsupp} \eqen This is further motivated by the phenomenological observation that the result is not very sensitive to the exact value of the parameter $\rho$.

\begin{figure}[tb]
  \hbox{
     \vbox{
	\begin{center}
	\mbox{
	\psfig{figure=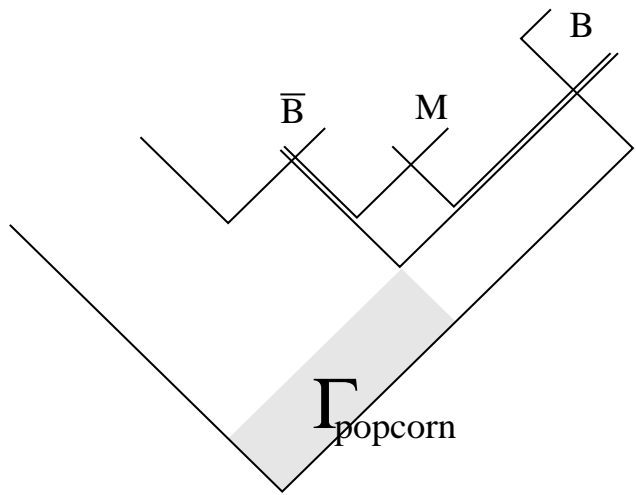,width=10cm,height=14cm}
	}
	\end{center}
    }
  }
  \vspace{-6 cm}
  \caption{\em The $\Gamma$ value for a popcorn system.}
  \label{f:Gpop}
\end{figure}

In the case of popcorn mesons, it is not obvious how to choose the corresponding effective $\Gamma$-value to be used in the suppression factor Eq~(\ref{e:Gsupp}). In our simulations we have used the area indicated in Fig~\ref{f:Gpop} which is the largest possible area when the curtain quarks do not exceed the speed of light. This assumption implies that these $\Gamma$-values are in general much smaller than corresponding values for diquark vertices without popcorn, which consequently influences the popcorn probability.

An important effect of the suppression for small $\Gamma$ is a suppression of rank 1 baryons (the effect being stronger for light baryons than for baryons with heavy c or b quarks). For low rank hadrons, $\left< \Gamma \right> $ is  smaller than the result in Eq~(\ref{e:meanG}), which is a limiting value in the centre. The average values $\left< \Gamma \right> $ for rank 1 and rank 3 baryons are shown in table~\ref{tab:gam}, and for the light baryons (represented by p and $\Lambda$) we see a significant difference. This suppression of low rank (light) baryons obviously results in softer momentum spectra, and the results will be further discussed in the following section.

\begin{figure}[tb]
  \hbox{
     \vbox{
	\begin{center}
	\mbox{
	\psfig{figure=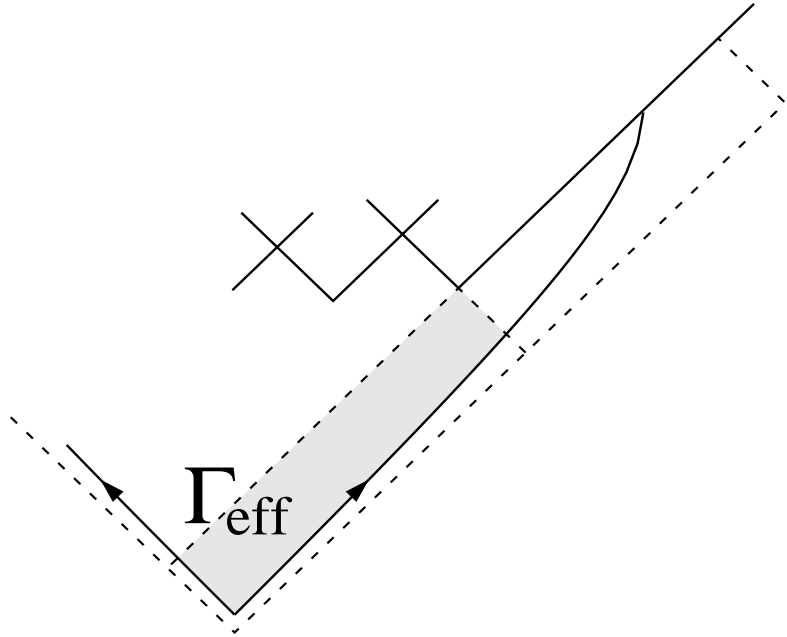,width=12cm,height=16.5cm}
	}
	\end{center}
    }
  }
  \vspace{-7 cm}
  \caption{\em In the case of heavy end quarks of mass $\mu_{\mrm q}$, the curtain pair production area $\Gamma$ is cut by the hyperbolic motion of the quarks, giving rise to a smaller $\Gamma_{eff} \approx \Gamma - \mu_{\mrm q}^2\ln(1/z)$.}
  \label{f:heavy}
\end{figure}

For heavy baryons with a c or a b quark, the large quark mass automatically implies larger values for $\Gamma$. This is the case also if we take into account the fact that the heavy quark moves along a hyperbola and not along the lightcone, as illustrated in Fig~\ref{f:heavy}. (Remember that in the string model these heavy hadrons can only be produced as first rank particles including an initial heavy quark.) Thus the whole area $\Gamma$ as defined in momentum space is not available for quark pair production but only the smaller area $\Gamma_{\mrm eff}$, marked in the figure. In table~\ref{tab:gam} we have for $\mrm \Lambda_c$ and $\mrm \Lambda_b$ presented corresponding average values, $\left< \Gamma_{\mrm eff} \right>$, and we note that these have approximately the same size as the $\left< \Gamma \right>$-values in the central region for light baryons. To obtain the values in table~\ref{tab:gam} we have been using a heavy quark fragmentation function which reproduces the experimental momentum spectra. This is the case for the fragmentation function suggested by Morris and Bowler~\cite{Bow}, which is obtained if the space-time area indicated in Fig~\ref{f:heavy} is interpreted as a ``colour coherence area'' in the Lund string model. This fragmentation function is in very good agreement with charm and bottom momentum spectra~\cite{jari}, and is the default option in JETSET version 7.4. Similar results are obtained if we use the Peterson fragmentation function~\cite{pet} fitted to the data.

At this stage, no new tunable parameter is introduced in the model. The values of both $\rho$ and $\delta a$ are determined by model predictions. However, the shift of the $a$-parameter is more complicated. The $\delta a$ value obtained above suppresses diquark vertices with large $z$ values, i.e.\ vertices close to the string ends. Thus it describes the very small probability for a curtain quark pair to be produced around such vertices. A similar suppression is expected also for ordinary \pair{q} production if the pair is not massless with zero transverse momentum~\cite{tunnel,jet}. The massive \pair{q} pair is produced somewhere along two hyperbolae around the vertex and the range of possible production points on these hyperbolae decreases when the vertex is close to the string ends. This effect is taken into account when tuning the $a$ parameter in the MC. Thus for baryon production there are two different effects which suppress small $\Gamma$-values, resulting in larger effective $a$-values. It is not at all obvious that the combined effect can be well described by simply adding the corresponding $\delta a$-shifts. For this reason we have in our MC (as in the original JETSET MC) left $\delta a$ as a tunable parameter, with expected best values in the range 0-0.5.

\section{Results} \label{sec-results}
We have implemented our model in a revised version of the JETSET Monte Carlo, and we will here discuss some results for $e^+e^-$ annihilation at the $Z^0$ energy. The results are also compared with default JETSET 7.4.

{\bf Flavour composition}\\
We have made a preliminary tuning of the flavour production parameters and in table~\ref{tab:mult} we present the particle multiplicities obtained for $\beta_{\mrm u}=0.6\mrm GeV^{-1}$, $\Delta\beta=1.2\mrm GeV^{-1}$, decuplet/octet$=0.19$ and $\delta a=0.5$. We note a very good agreement between our model and experimental data. The discrepancy is less than 1.5 standard deviations in all cases except for $\Sigma^\pm~(-1.7\sigma)$. Thus compared with JETSET default we get with fewer parameters an equally good or better agreement with data.

At high energies the effective suppression of first rank baryons does not affect noticeably the total multiplicity of light baryons. It does however reduce the production of heavy baryons containing a c- or a b-quark, which in the string model can only be produced as first rank particles. Compared to the default JETSET version these particles are reduced to about 70\% (For $\mrm \Lambda_c$ and $\mrm \Lambda_b$ the reduction is somewhat larger, around 60\%). Due to the large experimental uncertainties both the larger and the smaller results are consistent with data.

\begin{figure}[tb]
  \hbox{
     \vbox{
	\begin{center}
	\mbox{
	\psfig{figure=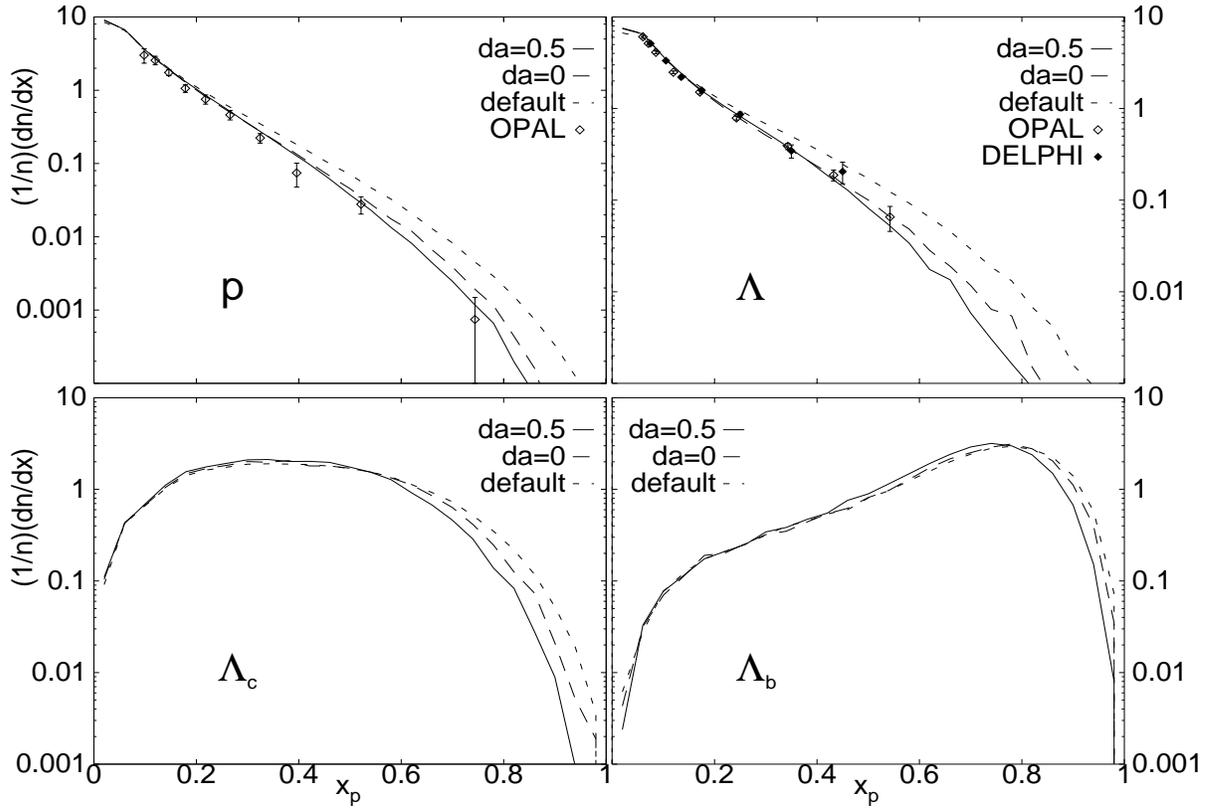,width=14cm,height=19cm}
	}
	\end{center}
    }
  }
  \vspace{-5cm}
  \caption{\em Momentum distribution $\frac{1}{n}\frac{dn}{dx_p}$ for {\em p}, $\Lambda$, $\Lambda_{\mrm c}$ and $\Lambda_{\mrm b}$. $x_p=p_{\mrm hadron}/p_{\mrm beam}$. Results from Monte Carlo simulations with $\Gamma$ suppression model are shown for two different values of $\delta a$, and compared to JETSET default and experiments. OPAL data are from refs~{\em \protect\cite{OPAL}}. DELPHI data are from ref~{\em \protect\cite{DELl}}.}
  \label{f:dist}
\end{figure}

{\bf Momentum distribution}\\
Baryons with large momenta are suppressed for two reasons. The most important one is the $\Gamma$-suppression in Eq~(\ref{e:Gsupp}). High momenta are also suppressed by a large value for the parameter $\delta a$. This suppression is, however, less effective. In Fig~\ref{f:dist} we show momentum distributions for protons and $\Lambda$-particles for two values of $\delta a$, 0 and 0.5. We note that a very good agreement with proton and $\Lambda$-data is obtained for both $\delta a$ values. In Fig~\ref{f:dist} we also show that the effect of $\Gamma$ suppression is rather small on $\Lambda_{\mrm c}$ and $\Lambda_{\mrm b}$ momentum distributions.

\begin{figure}[tb]
  \vspace{-2cm}
  \hbox{
     \vbox{
	\begin{center}
	\mbox{
	\psfig{figure=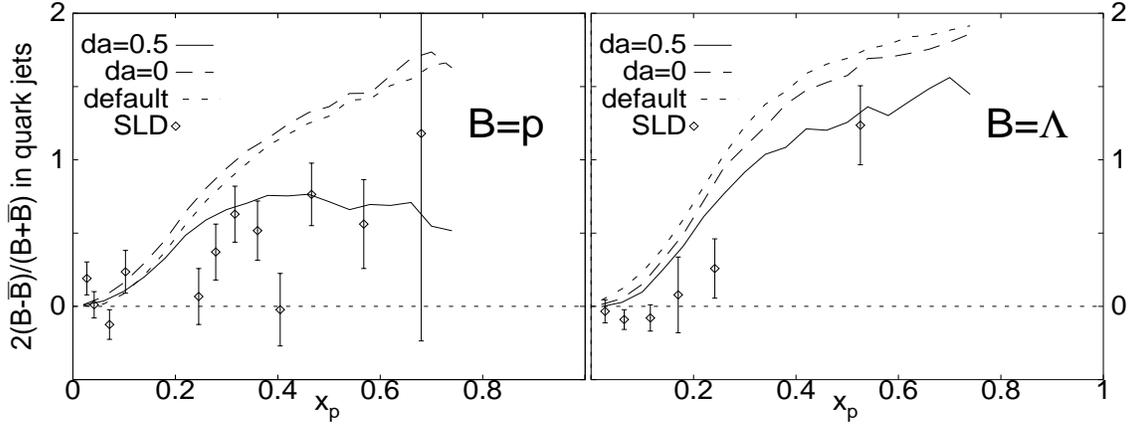,width=14cm,height=19cm}
	}
	\end{center}
    }
  }
  \vspace{-11 cm}
  \caption{\em The $x_p$ distribution of $\mrm 2*(B- \ol B)/(B+\ol B)$ for baryons and antibaryons in the same hemisphere as the outgoing quark. The model predicts more hard baryons than antibaryons in quark jets, as has been observed. Quantitatively, the predictions are sensitive to the $\delta a$ value. The plots show ${\rm p}$ and $\Lambda$ asymmetry for $\delta a=0$ and $0.5$, JETSET default and SLD preliminary data~{\em \protect\cite{SLD}}.}
  \label{f:SLD}
\end{figure}

Even if inclusive baryon spectra are fairly insensitive to the value of $\delta a$, this parameter can be restricted if we separate the quark and antiquark jets. In Fig~\ref{f:SLD} we show baryon-antibaryon asymmetries compared with data from SLD~\cite{SLD}. We note that data seem to favour a $\delta a$ value larger than 0, and that future studies with better statistics could be used to restrict the range of possible values for the $\delta a$ parameter. 

{\bf Correlations}\\
Correlation data can be used as a more thorough test of the model. As examples, multiplicity correlations for \pair{\Lambda} and $\Lambda$\anti{\Xi^-} pairs are also presented in table~\ref{tab:mult}. The number of correlated \pair{\Lambda} pairs (0.106 per event) is in fair agreement with data, and the number of $\Lambda$\anti{\Xi^-} pairs predicted to be 0.021 per event.

\begin{figure}[tb]
  \vspace{-2cm}
  \hbox{
     \vbox{
	\begin{center}
	\mbox{
	\psfig{figure=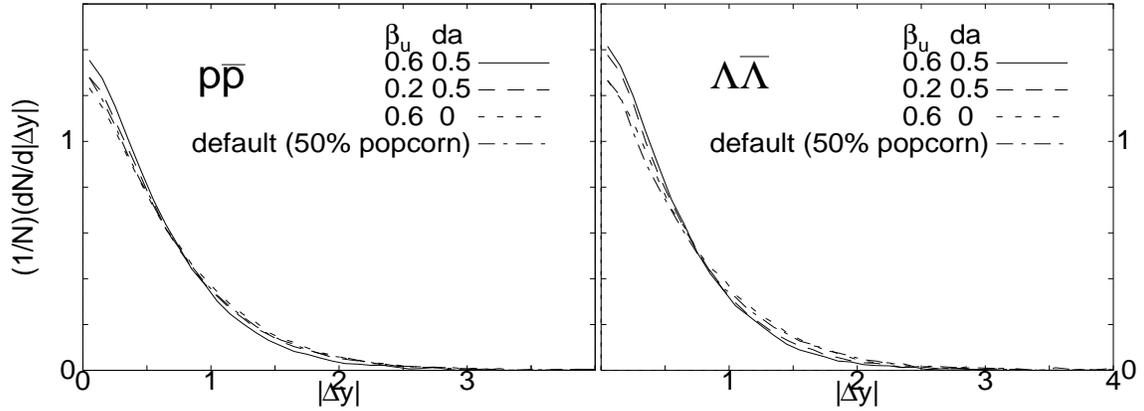,width=14cm,height=19cm}
	}
	\end{center}
    }
  }
  \vspace{-11cm}
  \caption{\em \pair{\Lambda} and \pair{p} rapidity correlations. $N\equiv N_{\mrm B \ol B}+N_{\mrm \ol B B}-N_{\mrm B B}-N_{\mrm \ol B \ol B}$. The curves corresponds, as indicated, to different values of $\beta_{\mrm u}$ and $\delta a$ and to the traditional JETSET MC with default parameters and 50\% popcorn.}
  \label{f:dy}
\end{figure}

Some results for rapidity correlations are presented in Fig~\ref{f:dy}, which shows distributions in $N \equiv N_{\mrm B\ol B'}-\frac{1}{2}(N_{\mrm BB'}+N_{\mrm \ol B\ol B'})$ vs.\ $|\Delta y|$. We have compared our results with default JETSET and not directly with experimental data, because it is difficult to implement exactly the same experimental cuts.  Experimental results, which have been compared with the JETSET MC, appear to be conflicting, as OPAL results favour a large popcorn parameter around 95\%~\cite{OPALy}, while DELPHI data agree with 50\% popcorn~\cite{DELl}. As seen in Fig~\ref{f:dy} there are only small differences between the theoretical curves, but we note that to some extent the result does depend on the value of $\delta a$. This is of interest because in published comparisons between experimental data and the JETSET MC only the popcorn parameter has been tuned.

As mentioned above correlations in ${\bar p}_\perp$ or azimuthal angle can be affected by the neglected small transverse momentum of the curtain quarks. For this reason we do not present predictions for these correlations. If wanted the $p_\perp$ of the curtain quarks could be implemented with one more parameter, which does not affect any other observable. One should note, though, that at higher energies, e.g.\ at LEP, correlations in ${\bar p}_\perp$ or azimuthal angle are mainly determined by jet production, while transverse momentum contributions from the fragmentation contributes very little.

{\bf Energy dependence}\\
Due to the suppression of first rank baryons one could expect that the energy dependence of the baryon production might be slightly changed. This could follow from the fact that at lower energies a larger fraction of the hadrons are first rank. The energy dependence of the baryon rate depends also on the amount of popcorn, and it turns out that for the parameter values presented above the energy variation between 10GeV and 91 GeV is not noticeably different from the original default JETSET MC.

We end by noting that our model contains two features which can  be treated separately in the MC, first the new scheme for flavour compositions and second the suppression in Eq~(\ref{e:Gsupp}) leading mainly to a softer momentum spectrum. Thus in order to understand which features are essential for the result, it is e.g.\ possible to make simulations with the old model for flavour generation but include the suppression for small $\Gamma$. For heavy baryons this gives a smaller suppression than our full model, as the rate is reduced to $\sim$85\% of the standard JETSET value.

\section{Summary}
A revised version of the ``pop-corn'' model for baryon production is presented, in which the stepwise production of the quarks in a baryon is carried through consistently. As in the popcorn model it is assumed that a \pair{q} pair with ``wrong colour'' (i.e.\ not with the colour which makes the string break) can be produced as a quantum fluctuation. One or more secondary \pair{q} pairs can be produced within the fluctuation causing the string to break and producing a \pair{B} pair plus possibly also mesons. We assume here that these secondary \pair{q} pairs are produced in just the same way as ordinary meson-producing \pair{q} breakups in the string. In this way the number of free parameters can be reduced, and yet the experimental production ratios for different baryon species can be reproduced.

Describing the initial ``wrong colour'' \pair{q} pair with a semiclassical model, we find a suppression of hard baryons, in particular of hard first rank baryons. This implies that the model reproduces well experimental momentum distributions for p and $\Lambda$.

Finally a set of predictions for \pair{B} correlations are presented.

{\Large {\bf Acknowledgments}}\\
We want to thank Torbj\"orn Sj\"ostrand for many valuable discussions, and for his guidance into the JETSET Monte Carlo code.

\begin{table}[p]
\begin{center}
\begin{tabular}{||c|c|c|c|c|c||c||} 
  \hline
 & \multicolumn{5}{c||}{relative rate of popcorn systems} & mean number of\\
$\beta_{\mrm u}$ & \multicolumn{5}{c||}{with N primary mesons} & primary mesons in \\
($\mrm GeV^{-1}$) & ~N=0~ & ~N=1~ & ~N=2~ & ~N=3~ & ~N$>$3~ & a popcorn system\\
\hline
$1.0$ & 0.69 & 0.23 & 0.06 & 0.01 & 0.01 & 0.52\\
$0.6$ & 0.60 & 0.26 & 0.10 & 0.03 & 0.02 & 0.62\\
$0.2$ & 0.46 & 0.27 & 0.14 & 0.07 & 0.05 & 1.05\\
\hline
\end{tabular}
\end{center}
\caption{\em The rate of popcorn systems with different number of popcorn mesons for the same tune as in table~\protect\ref{tab:mult} where $\beta_{\mrm u}$ is $0.6\mrm GeV^{-1}$, and similar tunes with $\beta_{\mrm u}=1\mrm GeV^{-1}$ and $0.2\mrm GeV^{-1}$.}
\label{tab:Npop}
\end{table}

\begin{table}[p]
\begin{center}
\begin{tabular}{||l|c|l|l||} 
  \hline
  & \multicolumn{3}{c||}{Multiplicity} \\
  & & \multicolumn{2}{c||}{JETSET7.4}\\
  hadron & experiments & default & modified\\
  \hline
p & 0.98$\pm$0.10 & 1.20 & 0.87 \\
$\Lambda$ & 0.371$\pm$0.015 & 0.387 & 0.363 \\
$\Sigma^0$ & 0.071$\pm$0.015 & 0.074 & 0.070 \\
$\Sigma^\pm$ & 0.177$\pm$0.023 & 0.141 & 0.137 \\
$\Xi^-$ & 0.0257$\pm$0.0014 & 0.0271 & 0.0274 \\
$\Delta^{++}$ & 0.124$\pm$0.065 & 0.189 & 0.118\\
$\Sigma(1385)^{\pm}$ & 0.044$\pm$0.008 & 0.074 & 0.056 \\
$\Xi(1530)^0$ & 0.0061$\pm$0.0011 & 0.0053 & 0.0062 \\
$\Omega^-$ & 0.0016$\pm$0.0005 &0.0007 & 0.0010 \\
  \hline
$\Lambda_{\mrm c}$ & 0.075$\pm$0.024 & 0.059 & 0.034 \\
All c baryons & & 0.083 & 0.056 \\
$\Lambda_{\mrm b}$ & & 0.033 & 0.020 \\
All b baryons & & 0.058 & 0.039 \\
  \hline
$\Lambda \ol \Lambda$ & 0.089$\pm$0.007 & 0.090 & 0.106 \\
$\Lambda \ol \Xi^- $ & & 0.019 & 0.021 \\
  \hline
\end{tabular}
\end{center}
\caption{\em Average multiplicities from JETSET 7.4 simulations of $1.6*10^6~e^+e^-$ events at $\mrm Z^0$ energies, compared to experimental data. The parameter values used in the new baryon production model are $\beta_u=0.6\mrm GeV^{-1},~\Delta \beta=1.2\mrm GeV^{-1}$, decuplet suppression $=0.19$ and $\delta a = 0.5$. Experimental data are from ref~{\em \protect\cite{revue} } $\Lambda_{\mrm c}$ data are from ref~{\em \protect\cite{Lam_c} }. No extensive tune of the meson parameters have been made.}
\label{tab:mult}
\end{table}

\begin{table}[tb]
\begin{center}
\begin{tabular}{||l|l|l||} 
  \hline
  hadron & rank 1 & rank 3 \\  \hline
  p & 1.9 & 2.9 \\
  $\Lambda_s$ & 1.9 & 2.9 \\ \hline
  $\Lambda_c$ & 2.6 & (only created as rank 1) \\
  $\Lambda_b$ & 2.7 & (only created as rank 1) \\
  \hline
\end{tabular}
\end{center}
\caption{\em $\left< \Gamma \right>$ for light baryons and $\left< \Gamma_{eff} \right>$ for heavy baryons. JETSET 7.4 default simulation of $e^+e^-$ collisions at $\mrm Z^0$ energies (using Bowler correction to $f(z)$ for heavy flavours).}
\label{tab:gam}
\end{table}


\end{document}